\documentstyle[aps,preprint,epsf,floats]{revtex} 
\tighten 
\begin{document} 
 
\title{The logic of causally closed space-time subsets} 
\author{H. Casini \\ 
{\sl Centre de Physique Th\'{e}orique, Campus de Luminy, F-13288 Marseille, 
France} \\ 
e-mail: casini@cpt.univ-mrs.fr} 
\maketitle 
 
\bigskip 
   
\begin{abstract} 
The causal structure of space-time offers a natural notion of an opposite or 
orthogonal in the logical sense, where the opposite of a set is formed by all 
points non time-like related with it. We show that for a general space-time 
the algebra of subsets that arises from this negation operation is a complete 
orthomodular lattice, and thus has several of the properties characterizing
 the algebra of physical propositions in quantum mechanics.  We think this 
lattice could be 
used to investigate the causal structure in 
an algebraic context. As a first step in this direction we show that 
 the causal lattice is in addition atomic, find its atoms, and give necessary  
and sufficient conditions for ireducibility.  \end{abstract} 
\bigskip
\section{Introduction} 
 
The logical propositions that represent physical questions 
in quantum mechanics are of the 
form `the state of the system is an eigenvector of the 
 operator $\cal{O}$ with eigenvalue $\lambda$', for an observable $\cal{O}$. 
The non 
commutativity of the observable operators gives place to a non Boolean 
logic for the propositions, in contrast to the case of the propositions 
about the phase space of a system in classical physics.  This is 
mathematically described by the orthomodular lattice structure of projectors 
in Hilbert space, which is often called quantum logic \cite{ql,beltram}. 
 
The algebra of propositions about the space in classical physics is Boolean. 
Consider for example 
a particle, and the set of propositions of the form `the particle is in the 
set $S$'. 
The opposite in this logic to the 
proposition given by the subset $S$ of the space is given by $-S$, the set of 
all points that do not belong to 
$S$.  A positive answer to $S$ implies a negative one to $-S$. The 
intersection and union of sets corresponds to the conjunction and disjunction 
of logical propositions about the physical system.  The algebra of 
propositions becomes the same as the Boolean algebra of subsets of the space 
in set theory. This structure is inherent to the space, because it will be 
the same for any system. 
 
However, as we will see in the following, the situation when including time 
is different. In the 
relativistic context the kinematics imposed by the causal structure will 
induce a non Boolean character for the algebra of propositions given by 
certain space-time subsets (see \cite{galileo} for the case of Galilean 
physics). A 
particle can not be concentrated at a given point $p$ in space-time because 
it existed in the past and will exist in the future of $p$, but we can think 
that the answer to the proposition given by a point is true if the particle 
 passes through $p$ and false if not. The causal structure gives a definite 
prescription of what the opposite means in this case, because if the particle 
passes through $p$ it can not pass through any point spatially separated from 
$p$. All the points non related by a time-like curve to a given set $S$ will 
form the causal and logical complement of $S$. 
 
A surprising fact about the algebra of propositions that arises along this 
line of thought from the causal structure of space-time alone, independently 
of the propagating physical system, is that it is also an orthomodular 
lattice. Though, we think, not widely known, this very interesting point was 
shown for Minkowski space some time ago in ref.\cite{cegla}. 
 
In this work we will show how to define the lattice of causally closed sets $ 
{\cal L}({\cal M})$ for a general space-time ${\cal M}$, even non causally 
well behaved, and prove that it is a complete orthomodular lattice. In 
addition, we will show that the lattice is atomic, and find its atoms.
 They are the 
 points for causally well behaved space-times, but in general they can be of 
three different kinds. We will also find the necessary 
and sufficient conditions for this lattice to be irreducible. 
 
The lattice ${\cal L}({\cal M})$ can have interest in two senses. We think 
that this lattice structure can be more fundamental than the causal 
structure of the space-time, much in the spirit of the studies of the 
postulates of quantum mechanics from the lattice perspective 
\cite{beltram}. In this sense this paper can be consider as a step in 
the study of the causal structure of space-times from an algebraic point of 
view. In fact, we find that certain undesirable aspects of the causal 
structure of general space-times as closed time-like curves can be 
eliminated or localized in the structure of ${\cal L}({\cal M})$, this 
lattice being largely independent of the details of the metric in the bad 
sectors of space-time. Two space-times related by a conformal transformation 
will have the same causal structure and lattice ${\cal L}({\cal M})$, 
however the converse can be false. 
 
The other aspect of interest is the possible relation of ${\cal L}({\cal M})$ 
with quantum mechanics and quantum field theory, motivated by 
the intriguing fact that causal structure and Hilbert space share an 
important piece of its mathematical structure. We will not develop in this 
sense here. We note, however, that the causally closed sets in ${\cal L}( 
{\cal M})$ appear naturally in the algebraic approach to quantum field 
theory. In that context, we have a set of $C^{*}$-algebras associated to 
space-time sets. To a set in ${\cal M}$ and its causal complement correspond 
two $C^{*}$-algebras that commute. 
However, the whole relation between the lattice 
structure and the operations among local algebras seems not to be completely 
clear at the 
moment \cite{haag}. We will say a bit more in Section III. 
 
 To make this article as much self contained as possible, in the 
 following Section we include a brief introduction to orthomodular 
lattices and some relevant mathematical preliminaries. Further details can be 
found in refs. \cite{beltram,bir,orthomodular}. In Section III 
we define the orthocomplemented complete lattice ${\cal L}({\cal M})$ and 
discuss several related possibilities for constructing causal set lattices. 
In Section IV we show that ${\cal L} ({\cal M})$ is orthomodular.  In 
Section V we show that it is atomic and find the atoms, which are an 
important piece in the understanding of the general structure of the lattice 
that will allows us to characterize irreducibility. In the last Section we 
make some comments and discuss directions open to future work. 
 
\section{Orthomodular lattices} 
 
The family of subsets ${\cal B}(U)$ of any set $U$ forms a Boolean algebra 
when considered with the relation of inclusion, and the operations of 
intersection, union, and complement (denoted $\subseteq$, $\cap$, $\cup$ and 
$-$ respectively). This is at the base of the relation between set theory 
and classical logic, where the propositions can be restated in terms of the 
subsets of a given set $U$ in set theory. To a given proposition ${\cal P}$ 
we assign a subset $S$ with the interpretation that an element belongs to $S$ 
if and only if it satisfies the proposition ${\cal P}$. Thus, the operations 
of implication, conjunction, disjunction, and negation between propositions 
are interpreted as the relation of inclusion, and the operations of 
intersection, union, and complement between subsets respectively. 
 
In quantum mechanics the space of individuals is given by a Hilbert space $ 
{\cal H}$. The corresponding propositions or questions that are answered 
with a yes or a no, are given by projection operators $P$, $P^{2}=P$. These 
operators must be observables, that is, auto-adjoint operators, $ 
P=P^{\dagger }$, such that their domain is all ${\cal H}$, and in consequence 
 they are 
also bounded. The projection operators are in one to one correspondence with 
the 
closed subspaces $V_{P}$ of ${\cal H}$, such that $P\left( \,V_{P}\right) =V$, 
 and $P(V_{P}^{\perp })=0$. Here we use the notation $W^{\perp }$ 
for the subspace orthogonal to $W$, which is always closed. 
 
The interpretation is that the proposition $P$ is true for the 
elements of $V_{P}$ and false for the elements of $V_{P}^{\perp }$. Here is 
 possible to talk in terms of propositions 
(projection operators) or in terms of closed linear subspaces indistinctly, 
as it happens in the case of classical logic, where it is possible to talk in 
terms of 
propositions or in terms of sets. 
 
The set ${\cal C}({\cal H})$ of closed subspaces of a Hilbert space can be 
furnished with a structure that resembles the one in ${\cal B}(U)$, where the 
relation of inclusion and the operation of intersection between linear 
subspaces are the set inclusion and intersection, but where the set 
complement is replaced by the operation of taking the orthogonal subspace, 
and the union of two sets by the sum of subspaces. The resulting structure 
 is called an orthomodular lattice. In what follows we will describe 
this structure in detail. 
 
The inclusion between subsets is an order relation, that 
makes ${\cal C} ({\cal H})$ (and ${\cal B} (U)$) a {\sl partially order set} 
(poset), 
 
\begin{eqnarray} 
A &\subseteq &A \,, \\ 
A &\subseteq &B\,,\,B\subseteq C\Rightarrow A\subseteq C\, , \\ 
A &\subseteq &B,\,B\subseteq A\Rightarrow A=B \,. 
\end{eqnarray} 
 
In ${\cal C}({\cal H})$ or ${\cal B}(U)$ there always exist elements that 
are the greatest lower bound (g.l.b.) and the lowest upper bound (l.u.b) of 
two given elements with respect to the order relation given by $\subseteq $. 
In ${\cal B}(U)$ they are given by the set intersection and union while in $ 
{\cal C}({\cal H})$ by the set intersection and sum of linear subspaces. We 
will denote generically the g.l.b. and l.u.b. of two elements $A$ and $B$ as 
$A\wedge B$, and $A\vee B$ respectively, and call them the meet and the join 
of $A$ and $B$. 
 
The meet and the join are associative and symmetric. A poset where 
the join and the meet of any two elements always exist is called a {\sl 
lattice}. A lattice is in addition {\sl complete} if the meet and the join 
exist for arbitrary families of elements. 
 
A lattice ${\cal L}$ have a {\sl unit} $I$ and a {\sl zero} element $O$ if 
$A\subseteq I$ and $O\subseteq A$ 
for every element $A\in {\cal L}$. 
 
 An element $A$ in a lattice has a 
{\sl complement} $B$ if it is $A\wedge B=O$ and $A\vee B=I$. 
The lattice is called {\sl orthocomplemented} (or {\sl ortholattice}) if 
there exist a unary operation $\perp :{\cal L} \rightarrow {\cal L}$ that 
assigns to every element $A$ a complement $A^{\perp}$, and in addition 
\begin{eqnarray} 
A &=&(A^{\perp})^{\perp} \,,  \label{negacion} \\ 
A &\subseteq &B\,\Rightarrow B^{\perp}\subseteq A^{\perp}\, . 
\end{eqnarray} 
 
In an orthocomplemented lattice it can be deduced that $(A\vee B)^{\perp 
}=A^{\perp }\wedge B^{\perp }$ and $(A\wedge B)^{\perp }=A^{\perp }\vee 
B^{\perp }$, what shows that the negation is a 
dual map with respect to the operations $ \wedge $ and $\vee $. 
 
Both, ${\cal B}(U)$ and ${\cal C}({\cal H})$ are complete and 
orthocomplemented. However, for the lattice ${\cal B}(U)$ holds the 
distributivity law 
\begin{equation} 
A\wedge (B\vee C)=(A\wedge B)\vee (A\wedge C)\,,  \label{distributivity} 
\end{equation} 
while it is easy to see that it does not hold for ${\cal C}({\cal H})$. An 
orthocomplemented distributive lattice is called a {\sl Boolean algebra}. 
However, in 
the lattice ${\cal C}({\cal H})$ holds a relaxation of the 
distributivity, called {\sl orthomodularity} 
\begin{equation} 
B\subseteq A\,,\,C\subseteq A^{\perp }\Rightarrow \,A\wedge (B\vee 
C)=(A\wedge B)\vee (A\wedge C)\,.  \label{ortho} 
\end{equation} 
Orthomodularity is equivalent to the simpler relation 
\begin{equation} 
B\subseteq A\Rightarrow \,A\wedge (B\vee A^{\perp })=B\,.  \label{or} 
\end{equation} 
For finite dimensional Hilbert spaces a stronger form holds, the 
{\sl modular} law, 
\begin{equation} 
B\subseteq A\,\Rightarrow \,A\wedge (B\vee C)=(A\wedge B)\vee (A\wedge C)\,, 
\end{equation} 
whose definition, as the one of distributivity, makes no use of the 
orthocomplementation operation. 
Distributivity implies modularity and, for an orthocomplemented lattice, 
modularity implies orthomodularity. 
 
We see that the orthomodular law 
(\ref{ortho}) means that distributivity holds under special circumstances 
for the elements involved. In the lattice of closed linear subspaces of the 
Hilbert space, ${\cal C}({\cal H})$, when using the representation of 
projection operators, these conditions can be rewritten as that the projector 
$P_{A}$ commutes with $P_{B}$ and $P_{C}$. In a general orthocomplemented 
lattice two elements $A$ and $B$ are said to {\sl commute }if 
\begin{equation} A{\sl =(}A\wedge B{\sl )\vee (A\wedge B}^{\perp }{\sl )\,}. 
\label{commu} \end{equation} The relation of commutativity is symmetric in 
orthomodular lattices, and the Boolean subalgebras are formed by mutually 
commuting elements. 
 
We will need a few additional definitions. With two lattices ${\cal L}_{1}$ 
and ${\cal L}_{2}$ we can construct the {\sl direct product} of lattices 
${\cal L}_{1}\times {\cal L}_{2}$ defined in the Cartesian product of sets, 
and where the operations are done component by component. The product of 
orthomodular lattices is orthomodular.  A lattice is {\sl irreducible} if it 
can not be written as a direct product of lattices. An orthocomplemented 
lattice is irreducible if and only if its {\sl centre}, that is, the set of 
elements 
that commute with all other elements, is equal to $\{I,\,O\}$. 
 
An {\sl atom} of a lattice is a non zero element $A$ such that $A$ and $O$ 
are the only elements included in $A$. A lattice is {\sl atomic} if every non 
zero element contains an atom. An orthomodular atomic lattice is also {\sl 
atomistic}, that is, every non zero element is the join of the atoms that 
contains. Both, $ {\cal B}(U)$ and ${\cal C}({\cal H})$ are atomic, its atoms 
being the points in $U$ and the unidimensional vector spaces in ${\cal H}$. 
 
\subsection*{Closure and Galois connection} 
 
We will now describe a fundamental method for constructing complete 
orthocomplemented lattices that will be useful later \cite{bir}. The 
orthocomplemented lattice will be constructed with a subset (not a 
sublattice) of a given complete lattice ${\cal L}$ (not necessarily 
orthocomplemented). The idea is to start with a would 
be complement operation $\perp $ and then to require the elements in the new 
lattice to be the $A\in {\cal L}$ that satisfy $\perp \perp A=A$. This  
is imposed by eq.(\ref {negacion}). This is similar to the case of a Hilbert 
space, where, if we start with ${\cal B}({\cal H})$ and propose the 
orthogonal to be the orthocomplement operation, the resulting lattice  is 
${\cal C}({\cal H})$, because the orthogonal of every set of vectors is a 
closed linear subspace. 
 
Given a complete lattice ${\cal L}$ a {\sl Galois connection} in ${\cal L}$ 
is a unary operation $g$ in ${\cal L}$ such that 
\begin{eqnarray} 
X &\subseteq &Y\Rightarrow Y^{g}\subseteq X^{g}\,,  \label{g1} \\ 
X &\subseteq &X^{gg}\,.  \label{g2} 
\end{eqnarray} 
The operation $c=g\circ g$ is called a {\sl closure operation}, and the 
elements $A$ of ${\cal L}$ that satisfy $A=A^{c}$ are called {\sl closed} 
with respect to the closure operation $c$. The closure $ 
c$ has the following properties 
\begin{eqnarray} 
X &\subseteq &X^{c}\,,  \label{a} \\ 
X^{c} &=&X^{cc}\,,  \label{b} \\ 
X &\subseteq &Y\Rightarrow X^{c}\subseteq Y^{c}\,.  \label{c} 
\end{eqnarray} 
If we want the Galois connection $g$ to be an orthocomplementation we 
must pick up the elements of the lattice such that the closure operation is 
the identity operator in them, as is enforced by eq.(\ref{negacion}). Indeed, 
if we further have \begin{eqnarray} 
X\wedge X^{g} &=&O\,,  \label{x1} \\ O^{g} &=&I\,,  \label{x2} 
\end{eqnarray} the set of closed elements ${\cal L}^{c}$ of ${\cal L}$ with 
respect to $c$, forms a complete orthocomplemented lattice where the order 
and the meet are the order and the meet in ${\cal L}$, the complement is 
given by $g$ and the join is the join in ${\cal L}$ followed by the operation 
of closure $c$.

Finally, we add that there is a very natural way to construct a Galois 
connection in the lattice ${\cal B}(U)$. Let $R$ be a symmetric reflexive 
relation between points in the set $U$, with no further requirements. Thus $ 
pRq\Leftrightarrow qRp$ and $pRp$. The operation $\perp: X\rightarrow 
X^{\perp}$ given by \begin{equation} X^{\perp }=\{q\,/\,-(qRp)\,\,\,\,\forall 
p\in X\}\,,  \label{cual} \end{equation} that is, the set of all points not 
related by $R$ with no point of $X$, is a Galois connection that satisfies 
eqs.(\ref{g1}, \ref{g2}) and (\ref{x1}, \ref{x2}).

\section{Causal set lattices} 
 
The space-time we will consider is a (Hausdorff) manifold ${\cal M}$ of 
dimension $d\geq 2$ with a pseudometric tensor $g_{\mu \nu }$ with signature 
$(+,-,..,-)$, smooth and non singular everywhere. We do not ask the 
space-time ${\cal M}$ to be time orientable nor to satisfy any causality 
condition (for example to be free from closed time-like curves). For a 
review of causal structure see refs. \cite{wald,haw,einste}. Given a 
definite naming of the local light cones of a point $p$ as future and past 
light cones, its causal future and past, $I^{+}(p)$ and $I^{-}(p)$, are 
defined as the set of points that can be reached by time-like curves starting 
at $p$ with tangent vector in the future and past light cone of $p$ 
respectively. When referring to time-like curves we always mean 
differentiable curves with continuous tangent vector except a finite number 
of points where the tangent can have a discontinuity, but keeping always in 
the same local light cone. These discontinuities can always be removed by 
modifications in arbitrarily small neighborhoods. The past and future of $p$ 
are open sets. We will use the symbol $I^{+}(S)\cup I^{-}(S)$ for 
the open set $\bigcup\limits_{q\in S}(I^{+}(q)\cup I^{-}(q))$ that always 
makes sense, even in non time-orientable space-times.

\begin{figure}[tb] 
\centering 
\leavevmode 
\epsfysize=4cm \epsfbox{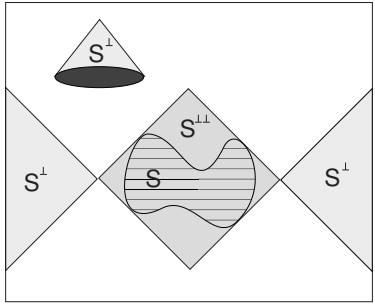} 
\smallskip 
\caption{The orthocomplement $S^{\perp }$, and the set generated 
$S^{\perp\perp } $ by a given set $S$. The space-time is the two-dimensional 
Minkowski space where we have taken out the closed set shown in black. Both, 
$S^{\perp }$ and $S^{\perp \perp }$ are in ${\cal L(M)}$, while $S$ is not. 
Note that in this example all points in the border of $S^{\perp}$ are in 
$S^{\perp}$ while all points in the border of $S^{\perp \perp}$, except the 
ones at the spatial corner, are in $S^{\perp \perp}$.} \end{figure} 
 
The natural orhtocomplement operation for sets in ${\cal M}$ based on 
the causal structure identifies all that is not time-like related to a 
given set with its complement. To be explicit, we can define in ${\cal M}$ a 
symmetric reflexive relation that relies only on the causal structure. Given 
two points $p$ and $q$ of ${\cal M}$ we write $p\sim q$, and say that $p$ and 
$q$ are time-connected or just connected, if there is time-like curve that 
passes through $p$ and $q$. Thus, $p\sim q$ is equivalent to $p\in 
I^{+}(q)\cup I^{-}(q)\cup \{q\}$. All points that are not time-connected to 
$p$ must be in the complement of $p$. Starting from this complement 
operation, as it was described in the previous Section, we can construct a 
complete orthocomplemented lattice coming from the Galois connection in 
${\cal B}( {\cal M})$ given by eq.(\ref{cual}). We call it the lattice of the 
causally closed sets ${\cal L}({\cal M})$. The orthocomplement is given by 
 
\begin{equation} 
S^{\perp }=\{p\,/\,p\text{ is not time-connected to any point of } 
S\}=-\left( S\cup I^{+}(S)\cup I^{-}(S)\right) \,, 
\end{equation} 
and, $S\in {\cal L}({\cal M})$ if and only if $S^{\perp \perp }=S$. The 
fig. 1 shows an example of $S^{\perp }$ and $S^{\perp \perp }$ for a set 
$S\subseteq {\cal M}$. 
An important property to keep in mind is that, for every set $S$, the 
opposite $S^{\perp }$ is already an element of ${\cal L}({\cal 
M})$, because $S^{\perp }=S^{\perp \perp \perp }$.  
 
A typical element $S$ of ${\cal 
L}({\cal M})$ in Minkowski space is a diamond shaped set, with the upper and 
lower cone being null surfaces included in $S$, while the points in 
the spatial corner may or may not be in $S$. However, in Minkowski space 
many other sets, including lower dimensional objects, are in the lattice. The 
bounded null surfaces of dimension $d-1$ or lower are sets in ${\cal L}({\cal 
M})$ while space-like surfaces of dimension $d-2$ or less are in ${\cal 
L}({\cal M})$, while spatial surfaces of dimension $d-1$ are not. On the 
other hand every set with at least two different points time-like connected 
will generate a set that contains an open set in ${\cal M}$. The sets 
generated by subsets in achronal surfaces form Boolean subalgebras. 
 
The meet in ${\cal L}({\cal M})$ is just the set intersection. The join of a 
family of 
elements is given by the set union followed by the causal 
closure (the double orthogonal) 
\begin{equation} 
\bigvee_{x\in F} A_x =\left(\bigcup_{x\in F} A_x\right)^{\perp \perp} \,. 
\end{equation} 
 The join of two sets is exemplified in fig. 2. For two spatially separated 
sets $A$ and $B$ the set $ A\vee B$ will be just the set union $A\cup B$, 
while otherwise $A\vee B$ will contain also at least all the time-like curves 
that connect points from $A$ to $B$. In general, for a set $A$ in ${\cal M}$, 
the causal closure 
$A^{\perp \perp}$ is bigger than the domain of dependence of $A$, taking this 
latter as the set of points such that all past (future) inextendible 
time-like curves from the point intersects the given set \cite{haw} 
(see fig. 2). 
 
The lattice ${\cal L}({\cal M})$ is not distributive nor modular in the 
general case, as it is shown for Minkowski space in fig. 3(a) and fig.3(b), 
while it is orthomodular as will be shown in the next Section and is 
illustrated in fig. 3(c).

As mentioned in the Introduction, the lattice ${\cal L}({\cal 
M})$ has a logical interpretation in terms of propositions for classical 
particles. The proposition corresponding to a space-time subset $S$ is given 
by ``the particle passes through S''. However, for an arbitrarily chosen  
set $S$ the 
logical opposite of that proposition is ``the particle does not passes 
through $S$''. In general this is not of the form ``the particle passes 
through $T$'' for some space-time subset $T$. The prescription of taking 
causally closed sets
 is just what is needed to have a closed algebra of propositions coming from 
space-time subsets in the above sense. In contrast to the case of the Boolean 
logic of space subsets, here only some subsets of space-time have an opposite 
in terms of space-time subsets, and the resulting algebra is non Boolean. 
The logical interpretation of the operation $\wedge$ between two propositions 
corresponding to the sets $A$ and $B$ is given by the maximal space-time 
proposition 
that 
implies ``the particle passes through $A$ and $B$'', while the proposition 
corresponding to $A\vee B$ is given by the minimal space-time proposition 
implied by ``
the particle passes through $A$ or $B$''. 
 
It is immediate that when the manifold ${\cal M}$ is non 
connected the lattice ${\cal L}({\cal M})$ is the direct product of the 
lattices ${\cal L}({\cal M}_{i})$ of each connected component ${\cal M}_{i}$ 
of ${\cal M}$. In this 
case the lattice is reducible. We will see a sufficient condition for 
irreducibility in Section V. 
 
\begin{figure}[tb] 
\centering 
\leavevmode 
\epsfysize=4cm \epsfbox{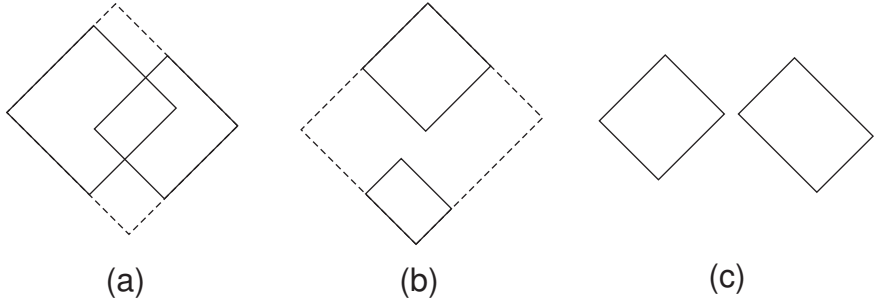} 
\smallskip \caption{Three cases for the join of two sets in two-dimensional 
Minkowski space. The set formed by the join is shown with dashed lines. In 
case (a) the join is equal to the domain of dependence of the set union, 
while in case (b) it is not. The case (c) shows the join of two spatially 
separated sets, that coincides with the set union. } \end{figure} 
 
One can try to construct other lattices based on the causal structure of $ 
{\cal M}$. For example, the basic symmetric and reflexive relation between 
points used to define the orthocomplement can be taken as the relation $pSq$, 
that holds if there is a space-like curve passing through $p$ and $q$, or 
$pCq$, 
that holds if there is a causal curve (i.e. a curve with time-like or null 
tangent) passing through $p$ and $q$, or $pNq$, that holds if there is null 
curve passing through $p$ and $q$, or several logical combinations between 
them (no discontinuities are allowed in the tangent to null and causal 
curves).  Accordingly, three orthocomplemented and complete lattices can be 
constructed, ${\cal L}_{S}({\cal M})$, ${\cal L}_{C}({\cal M})$ and ${\cal 
L}_{N}({\cal M})$ respectively. The lattice ${\cal L}_{S}({\cal M})$ is 
trivial when $d\geq 3$ because a space-like curve 
from a point can get inside the light cone, while in two dimensions it 
does not add anything new to $ {\cal L}({\cal M})$. The same 
triviality will occur in ${\cal L}({\cal M})$ if the metric would have 
more than one time directions. It is easy to see that ${\cal L} 
_{C}({\cal M})$ and ${\cal L}_{N}({\cal M})$ fail to be orthomodular in the 
simplest case of Minkowski space. It seems that in the case of 
${\cal L}_{N}({\cal M})$ it is not possible to change slightly the definition 
of the lattice to make it 
orthomodular. The lattice ${\cal L}_{C}({\cal M})$ differs from ${\cal L}( 
{\cal M})$ fundamentally in details of the borders of the sets, what 
 is crucial for orthomodularity. Other logical combinations seem to 
lead to the same lattices or to uninteresting and trivial cases. 
 
\begin{figure}[tb] 
\centering 
\leavevmode 
\epsfysize=4cm \epsfbox{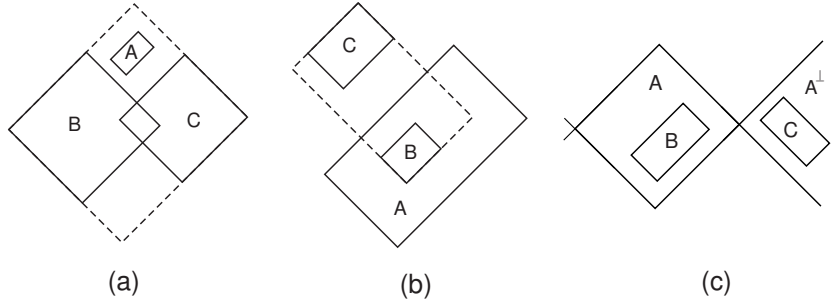} 
\smallskip \caption{Examples in two dimensional minkowski space. (a) 
Distributivity does not hold. We see that $A\wedge B=A\wedge C=O$, but 
$A\wedge (B\vee C)=A $. (b) The modular law does not hold. It is $B\subseteq 
A$, however $ A\wedge (B\vee C)$ is greater than $(A\wedge B)\vee (A\wedge 
C)=B$. (c) Orthomodularity holds. Under the conditions $B\subseteq A$, $C$ 
$\subseteq A^{\perp }$, the distributivity $A\wedge (B\vee C)=(A\wedge B)\vee 
(A\wedge C)=B$ holds.} 
\end{figure} 
 
One may wonder if there is another way for constructing a causal lattice 
that yielding open sets or sets with better topological properties. A 
lattice of open sets can be constructed using the complete lattice 
${\cal O}({\cal M})$ 
of open sets in ${\cal M}$ as a base where to define the Galois connection 
\cite{bir}. The join in ${\cal 
O}({\cal M})$ is the usual set union and the meet of arbitrary families of 
open sets is given by the set intersection followed by the operation of 
interior. Then we define $ S^{\perp }$ for an open $S$ as 
\begin{equation} 
S^{\perp}=-(\overline{I^{+}(S)\cup I^{-}(S)})\,, 
\end{equation} 
where here $\overline{A}$ 
means the usual topological closure of $A$. This satisfies eqs.(\ref {g1}, 
\ref{g2}) 
and (\ref{x1}, \ref{x2}), and lead to a complete orthocomplemented lattice 
${\cal L}_{{\cal O}}({\cal M})$ of open sets.  However, again due to details 
in the borders of the sets, the lattice is not orthomodular, as is shown in 
fig. 4. It can be shown, at least for globally hyperbolic 
space-times, that distributivity in ${\cal L}_{{\cal O}} ({\cal M})$ holds 
under a little more stringent conditions than the required by 
orthomodularity. In fact, instead of eq.(\ref{or}) we have 
\begin{equation} \label{apro} \overline{B}\subseteq A\Rightarrow \,A\wedge 
(B\vee A^{\perp })=B\,.  \end{equation} 
 
It seems essential for orthomodularity that the lattice should contain 
along with the diamond shaped sets at least the null surfaces. Once we have 
null surfaces in 
the lattice its intersection will generate lower dimensional sets as lines 
and points. 
 
This makes difficult a correspondence between the elements in an 
orthomodular lattice of causal sets with $C^{*}$-algebras in the algebraic 
approach to quantum field theory, because the points should be assigned non 
trivial algebras (two points can generate a set that includes an open set) 
(see the discussion in \cite{haag}). The lattice ${\cal L}_{{\cal O}}({\cal 
 M})$ would be a better candidate, but it is only approximately orthomodular 
 in the sense of eq.(\ref{apro}) (weaker conditions for distributivity can 
also be found). 
 
Form now on we will only refer to ${\cal L}({\cal M})$. From its definition 
we see that it does not change with conformal transformations, and thus it is 
a 
property of the conformal structure. 
 
\begin{figure}[t] 
\centering 
\leavevmode 
\epsfysize=4cm 
 \epsfbox{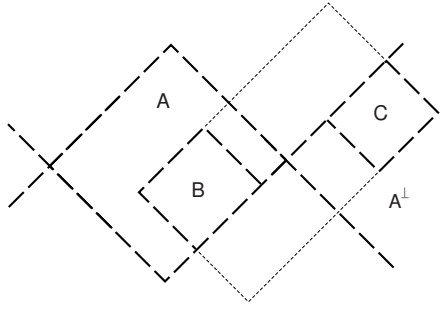} 
\smallskip 
\caption{Orthomodularity does not hold for ${\cal L}_{{\cal O}}({\cal M})$. 
All sets are open. The sets $B\subseteq A$ and $C\subseteq A^{\perp }$ share 
sectors of the border with $A$ and $A^{\perp }$ respectively. The set $ 
B\vee C$ is shown with short dashed lines. The lack of orthomodularity can be 
traced back to the absence of the null surface joining $B$ and $C$ in the 
orthogonal sets $B^{\perp}$ and $C^{\perp}$} \end{figure} 
 
\section{Orthomodularity of ${\cal L}({\cal M})$} 
 
We will now prove that ${\cal L}({\cal M})$ is orthomodular in any 
space-time. First, we will prove the following previous result. 
 
\bigskip 
 
{\bf Lemma 1}: Given a time-like curve $\gamma (x)$ parametrized by a real 
variable $x$ in a closed interval $x\in [a,b]$ and a set $S\in {\cal L}( 
{\cal M})$, the intersection $\gamma([a,b]) \cap S$ is empty or it is a closed 
segment of $\gamma $, corresponding to $x\in [c,d]\subseteq [a,b]$. 
 
\smallskip 
 
Suppose $x_{1}\in [a,b]$ and $x_{2}\in [a,b]$, with $x_{1}\le x_{2}$ are 
such $\gamma (x_{1})\in S$ and $\gamma (x_{2})\in S$, then it is $ 
\gamma ([x_{1},x_{2}])\subseteq S$. Otherwise, if $y\in [x_{1},x_{2}]$ is 
such that $\gamma (y)$ is not in $S$, it can be connected by a time-like 
curve to 
$ S^{\perp }$ (if not it would be in $(S^{\perp })^{\perp }=S$), and thus 
either $\gamma(x_{1})$ or $\gamma(x_{2})$ is connected by a time-like curve 
to $S^{\perp }$, contradicting the assumptions. Then, if the intersection 
$\gamma([a,b]) \cap S$ is non empty, let $c$ and $d$ be the infimum and 
supremum of 
the points $x$ in $ [a,b]$ such that $\gamma (x)\in S$. The point $\gamma 
(c)$ is connected with $S$ so $\gamma (c)\notin $ $S^{\perp }$, then if 
$\gamma (c)\notin S$ it is connected also with $S^{\perp }$, and thus belong 
to the open set $ I^{+}(S^{\perp })\cup I^{-}(S^{\perp })$. A neighborhood 
of $\gamma (c)$ will belong to $I^{+}(S^{\perp })\cup I^{-}(S^{\perp })$ and 
will not be in $ S$, what is not possible, so $\gamma (c)\in S$. Similarly 
we have $\gamma (d)\in S $. 
 
\bigskip 
 
Therefore, a set $S$ in ${\cal L}({\cal M})$ contains all time-like segments 
between points in $S$ and also the time-like border of $S$, that is, the 
border of $S$ accessible from $S$ by time-like curves. The points in the 
space-like border of $S$ may or may not belong to $S$. Using Lemma 1 we prove 
 
\bigskip 
 
{\bf Theorem 1: } ${\cal L}({\cal M})$ is orthomodular 
 
\smallskip 
 
Orthomodularity is the implication $B\subseteq A\Rightarrow A\wedge (B\vee 
A^{\perp })=B$ for any $A$ and $B$ in ${\cal L}({\cal M})$. If the set $B$ is 
included in $A$ it is also included in $A\wedge (B\vee A^{\perp })$. Then we 
have to show that $B\subseteq A\Rightarrow A\wedge (B\vee 
A^{\perp })\subseteq B$. To start we assume that $B\subseteq A$. 
Let $p\in A\wedge (B\vee A^{\perp})$, so $p\in A$ and $p\in 
B\vee A^{\perp }$. Therefore no time-like curve connects $p$ with $A^{\perp 
}$ nor with $B^{\perp }\wedge A$. We have to prove 
that $p\in B$. Then, let us suppose that $ 
p\notin B$ and show that it leads to a contradiction.  This assumption implies 
that there is a time-like 
curve $\gamma $ that connects $p$ with a 
point $ 
q\in B^{\perp }$. As $p$ is not connected with $A^{\perp}$ nor with $B^{\perp}
\wedge A$ it is  $q\notin A$ and $q\notin A^{\perp }$. 
A segment of $\gamma 
$ with end points $p$ and $q$ intersects $B^{\perp }$ and $A$ in non empty 
sets, which according to Lemma 1 are closed segments. These are disjoint, 
because otherwise $p$ 
could be connected with a point in $B^{\perp }\wedge A$. Thus, there is a 
point $r$ in the segment of $\gamma $ between $p$ and $q$ that is not in $ 
B^{\perp }$ nor in $A$. As  $r\notin B^{\perp}$, $r$ can be connected with a
 point $s\in B$. Thus, as $r$ is in the segment of $\gamma$ between $p$ and 
$q$ it is either $q$ connected with $s$ or $p$ connected with $s$. However, 
the point $q\in B^{\perp}$ can not be connected with $s\in B$. Then $p\in A$
 is connected with $s\in 
B\subseteq A$ by a curve passing through $r$. Then, by Lemma 1, it is 
$r\in A$, what shows 
a contradiction. Therefore $p\in B$ and orthomodularity holds. 
 
\bigskip 
 
Of course, there are orthomodular lattices that are not of the form ${\cal L} 
({\cal M})$ for a space-time ${\cal M}$. In the following Section we go a 
step further in the characterization of causal structures from the lattice 
theoretical point of view by showing that ${\cal L}({\cal M})$ is atomic, 
and finding its atoms. In the remaining part of this Section we will extract 
what are the essential properties of the causal structure used in the proof 
of orthomodularity (see ref.\cite{pp} for a related discussion). 
 
For constructing an orthocomplemented lattice we used a symmetric 
reflexive relation between points in the space-time with no further 
properties. Now we can read from the previous theorem 
what  additional structure of the causal relation $\sim $ that we used in 
the definition of ${\cal L}({\cal M})$ is central in the proof of 
orthomodularity. Basically, what we need is a set of curves in a set ${\cal M 
}$, locally one to one functions of (possible infinite) open intervals of 
real numbers to ${\cal M}$, that we can call time-like curves. Given a 
 point $p$ and any curve $\gamma $ with $\gamma (x)=p$, the points $\gamma 
(y)$ with $y\neq x$ belong to two sets, non necessarily disjoint, that we can 
call the past and future sets of $ p$. We have either all $\gamma (y)$ for 
$y>x$ belong to the future of $p$ and all $\gamma (y)$ for $y<x$ belong to 
the past of $p$ or vice versa. If a point $q$ is in the future set of $p$, 
then either all the future set of $q$ or the past set of $q$ are included in 
 the future set of $p$, and, in addition, there is a neighborhood of $q$ (in 
the sense induced by the real numbers) in all time-like curves passing 
through $q$ that belongs to the future set of $p$. The same can be said 
regarding the past set of $p$. 
 
These geometrical properties, that are immediate for the 
time-like curves in a general space-time ${\cal M}$, are sufficient in order 
to have an orthomodular lattice generated by the relation $\sim $. We see 
that certain sort of transitivity condition is essential for the causal 
relation but, for non orientable space-times no global transitive relation 
can be defined. There is also a condition of continuity. None of these is 
respected by the relation $N$ defined in the preceding Section, while the 
relation $C$ does not respect the continuity condition. 
 
\section{Atoms} 
 
An atom $S$ in ${\cal L}({\cal M)}$ must be equal to the set $ 
p^{\perp \perp }$ generated by any point $p\in S$, because $p^{\perp \perp }$ 
is a non empty element of ${\cal L}({\cal M})$ included in $S$. Therefore, 
all atoms are of the form $p^{\perp \perp }$ for some point $p$, what shows 
what are the sets among which to look for atoms. However, the element $ 
p^{\perp \perp }$ need not be an atom. By contrast, if $p$ as a subset of $ 
{\cal M}$ is an element of ${\cal L}({\cal M)}$, $p^{\perp \perp }=p$, then 
it is an atom. This is the case for all points in Minkowski space. We 
will now analyze which of the points of ${\cal M}$ are atoms in the general 
case. 
 
As we noted before, if two points $p$ and $q$ belong to $S$ in ${\cal L}( 
{\cal M)}$, then all points in time-like curves connecting $p$ and $q$ are 
also in $S$. If there is a time-like segment $\gamma $ connecting a point $p$ 
with itself, then $p$ can not be an atom because $\gamma \subseteq p^{\perp 
\perp }$. There are two types of such curves, as shown in fig. 5. One, the 
{\sl closed time-like curves}, that are totally included in the intersection 
$I^{+}(p)\cap I^{-}(p)$, where a segment of the curve between $p$ and $p$ 
start and end with its tangent vector in the same light cone of $p$. The 
other type, that we will call {\sl vertex curves}, where a segment between $ 
p $ and $p$ start and end with tangent vectors at $p$ in different light 
cones of $p.$ In this later case we will call $p$ a {\sl vertex}. For the 
existence of these types of curves the manifold must not be simply connected 
\cite{einste}. In the vertex curve case we see that the space-time is not 
time orientable. However, there are time non orientable space-times without 
vertex curves. The vertex curves mark space-times where the notion of future 
and 
past have no sense for a single observer. The points $p$ that belong to 
closed time-like curves or are vertex points can be characterized as all 
points that belong to $I^{+}(q)\cap I^{-}(q)$ for some $q$ of ${\cal M}$ 
(any $q$ of $\gamma $ in the open segment between $p$ and $p$ for example). 
We will call it the set ${\cal B}$ of bad points of ${\cal M}$, ${\cal B}$ $ 
=\cup _{q\in {\cal M}}I^{+}(q)\cap I^{-}(q)$. As union of open sets, ${\cal B} 
$ is also open. 
 
The set of points in closed time-like curves ${\cal C}$, and 
the set of vertices ${\cal V}$, are also separately open, and by definition $ 
{\cal B=C\cup V}$. To see this, let $p\in \gamma$, where $\gamma$ is a 
closed time-like curve, and take a sufficiently small, open, time orientable, 
normal neighborhood $U$ of $p$ \cite{haw}. Let $q$ and $r$ be two points of 
an interval of $\gamma$ included in $U$ that contains $p$, 
respectively  in the future $I_{U}^{+}(p)$ and past $I_{U}^{-}(p)$ 
of $p$, where $I_{U}^{+}$ and $I_{U}^{-}$ are the notions of future and 
past in the submanifold $U$. Doing a composition with the curve 
$\gamma$, we have that all points in the open 
neighborhood $I_{U}^{-}(q)\cap I_{U}^{+}(r)$ of $p$ 
 are connected by a closed time-like curve to every point in $\gamma$. 
Thus, the set of points connected by closed time-like curves to a point $p$ 
 form an open subset ${\cal C}_p$ of ${\cal C}$. It is immediate that the 
 sets of the form ${\cal C}_p$ are 
either disjoint or identical, so ${\cal C}$ is the union of disjoint sets of 
the form ${\cal C}_p$. 
Note that for any time-like curve $\delta$ that passes through $p$ we have 
that there is an open interval around $p$ in $\delta$ that is included in a 
 closed 
time-like curve trough $p$. 
The proof that ${\cal V}$ is open can be done similarly. 
 
\begin{figure}[t] 
\centering 
\leavevmode 
\epsfysize=4cm \epsfbox{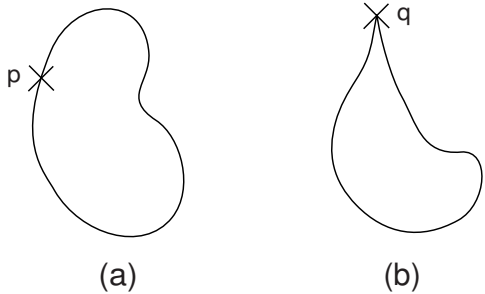} 
\smallskip 
\caption{(a) A closed time-like curve. All points in the curve are 
in the past and the future of all points in the curve. (b) A vertex curve 
with vertex point $q$. The vertex is in the past and the future of all points 
in the vertex curve (except $q$). } \end{figure} 
 
The following Lemma solves the question of when a point in ${\cal M}$ is an 
element, and therefore an atom, in ${\cal L}({\cal M})$ 
 
\bigskip 
 
{\bf Lemma 2: }the set formed by a point $p$ in ${\cal M}$ is an element and 
an atom in ${\cal L}({\cal M})$ if and only if $p\notin {\cal B}$ 
 
\smallskip 
 
As we have seen if $p\in {\cal B}$ the set formed by $p$ is not an 
element of ${\cal L}({\cal M})$ and can not be an atom. We will show that if 
$p$ is not an atom then $p\in {\cal B}$. Let us suppose $p^{\perp \perp 
}\neq p$. Thus, there is a point $q\in p^{\perp \perp }$ with $q\neq p$. The 
point $q$ must be time connected with $p$ otherwise it would be $q\in 
p^{\perp }$. Therefore $p\in I^{+}(q)\cup I^{-}(q)$ and a connected, time 
orientable, open neighborhood $U$ of $p$ is included in 
$I^{+}(q)\cup I^{-}(q)$. 
Thus, as every point in $U$ is connected with $q\in p^{\perp \perp }$, it 
must be that $U\cap p^{\perp }$ is empty. In other words $p^{\perp }$ is 
separated from $p$. Then all points $r$ in the open connected set $U-\{p\}$ 
are in $I^{+}(p)$ or in $I^{-}(p)$. With a given continuous time orientation 
in $U$ we have for every $r\in U-\{ p\}$ that $p$ is in $I^{+}(r)$ or in 
$I^{-}(r)$, and 
the sets $\{r\,/\,r\in (U-\{p\})\,,\,p\in I^{+}(r)\}\,$ and $\{r\,/\,r\in 
(U-\{p\})\,,\,p\in I^{-}(r)\}$ are open. They cover all the connected set $ 
U-\{p\}$, and thus must have a non empty intersection, with a point $s$. 
Therefore $p$ belongs to $I^{+}(s)\cap I^{-}(s)$ and thus $p\in {\cal B}$. 
 
\bigskip 
 
We have shown that in the closed set ${\cal M-B}$ all points are atoms. Next 
we will show that a different kind of atoms can exist in space-times that 
have both types of defects, closed time-like and vertex curves. Consider the 
time-like curve $\beta$ of fig. 6. We will call such a curve a {\sl bridge}. It is 
a closed time-like curve but also a past and future vertex curve for all 
points in the curve. The essential property of a bridge is that for any 
point $p$ in it one can construct a time-like curve that passes through $p$ 
with tangent in any light cone and passes through any point $q$ in the 
bridge with tangent in the any light cone of $q$. Thus any two points time 
connected with two points in the bridge $\beta $ are also connected to each 
other. Then we have the following 
 
\bigskip 
 
{\bf Lemma 3: }the set $\beta ^{\perp \perp }$ generated by any bridge $\beta 
$ is an atom 
 
\smallskip 
 
Let $p\in \beta $ (see fig. 6). If $q\in p^{\perp \perp }$ then $q$ is 
connected with $p$. The set $q^{\perp \perp }$ is included in $p^{\perp 
\perp }$ and if they are different it must be $p\notin q^{\perp \perp }$. 
In that case $p$ must be connected with $q^{\perp }$. But that is not possible 
because as $p$ belongs to a bridge it would lead to $q$ connected with $ 
q^{\perp }$. Thus $q^{\perp \perp }=p^{\perp \perp }$ for every $q\in 
p^{\perp \perp }$ and $p^{\perp \perp }$ is an atom that includes $\beta $. 
 
\bigskip 
 
\begin{figure}[tb] 
\centering 
\leavevmode 
\epsfysize=4cm \epsfbox{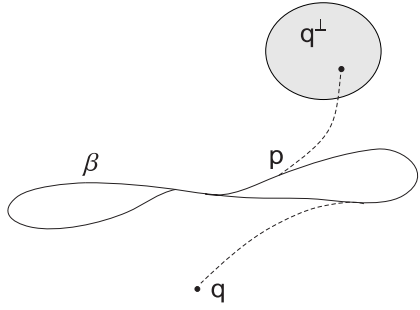} 
\smallskip 
\caption{A bridge time-like curve $\beta $. If a point $q$ is in $\beta 
^{\perp \perp } $ it must be connected with $\beta $. A point in $q^{\perp }$ 
can not be connected with a point $p$ in $\beta $, otherwise it would be 
connected with $q$. Thus, $\beta $ is in $q^{\perp \perp }$. Note that every 
two points in $\beta$ can be connected by a time-like curve (based on 
$\beta$) with tangent in any light cone of the points. This does not happen 
for simple closed time-like curves or vertex curves.} \end{figure} 
 
We will call the set $\beta^{\perp \perp}$ generated by a bridge $\beta$ a 
{\sl bridge atom}. Any bridge 
atom contains an open set. We also have that two atoms cannot 
intersect. Therefore in a paracompact manifold there can be at most a 
numerable amount of bridge atoms. The bridge atoms must be totally included 
in the set of bad points ${\cal B}$. 
 
We have shown what happens to the sets $p^{\perp \perp }$ for $p\in {\cal M-B 
}$ or $p$ in $\beta ^{\perp \perp }$ where $\beta $ is a bridge. In both 
cases $p^{\perp \perp }$ is an atom. We will see that this is not always the 
case. First consider the case where $p$ is a vertex. We have 
 
\bigskip 
 
{\bf Lemma 4:} The set $p^{\perp \perp }$ where $p$ is a vertex contains 
a point in ${\cal M-B}$ or a bridge atom 
 
\smallskip 
 
Let $\gamma (z)$, $z\in [0,1]$, and $\gamma (0)=\gamma (1)=p$ be a vertex 
curve of the vertex $p$. Let us suppose that every point in $\gamma $ is in $ 
{\cal B}$. For every value of the parameter $z\in [0,1]$ we define the future 
light cone 
of $\gamma (z)$ at $z$ as the light cone of $\gamma (z)$ given by the tangent 
vector of 
$\gamma$ at $z$. 
Thus, the direction of 
the tangent vector at the point $z=1$ will mark the future light cone of $p$ at $ 
z=1 $, but it will be a different light cone of the point $p$ than the future 
light cone 
at $z=0$. For any $z$ such that the point $q=\gamma (z)$ is a vertex, 
we can define it as a future or a past vertex according the vertex curve 
based at $q$ starts with tangent in the future or the past light cone of $q$ 
at $z$. 
  Let us call $W^{+}\subseteq [0,1]$ and $W^{-}\subseteq [0,1]$ the set of 
values 
of the parameter that have future and past vertices respectively. These sets 
are non empty as $0$ has a future vertex and $1$ has a past vertex. If $ 
w_{1} $ has a past vertex and $w_{2}$ has a future vertex and $w_{1}\leq 
w_{2}$, then both vertex curves together with $\gamma ([w_{1},w_{2}])$ will 
form a bridge. Let the infimum of $W^{-}$ be $ 
w^{-}$ and the supremum of $W^{+}$ be $w^{+}$, and assume $w^{+}\leq w^{-}$. 
The 
points in $\gamma ([w^{+},w^{-}])$ must belong to closed time-like 
curves, $\gamma ([w^{+},w^{-}])\subseteq {\cal C}$ (the points $w^{+}$ and 
$w^{-}$ 
 can not be vertices as ${\cal V}$ is open). But as ${\cal C}$ is 
open there must be $x^{+}$ and $x^{-}$ with $x^{+}<w^{+} \leq w^{-}<x^{-}$ and 
such $\gamma ([x^{+},x^{-}])\subseteq {\cal C}$. For each $y\in 
[x^{+},x^{-}] $ let $I_{\gamma (y)}$ be an open interval of $\gamma$ 
around the point $\gamma (y)$ that is included in a closed time-like curve 
through $\gamma (y)$. The sets $[x^{+},x^{-}]\cap 
\gamma ^{-1}(I_{\gamma (y)})$ are open in the topology of 
the interval, and cover the compact $[x^{+},x^{-}]$. After extracting a 
finite covering and gluing together a finite number of closed curves, we have 
 that $x^{+}$ and $x^{-}$ are connected by a closed time-like curve $ 
\delta $ that includes $\gamma ([x^{+},x^{-}])$. Then it is easy to see that 
the union 
 of $\delta-\gamma ([x^{+},x^{-}])$ with the 
vertices at $x^{+}$ and $x^{-}$ is a bridge. 
 
\bigskip

Then let us consider now the case of points in closed time-like curves. 
Let $p$ be a point in ${\cal C}$ and ${\cal C}_p$ be the connected component 
of ${\cal C}$ that includes $p$. It is  ${\cal C}_p\subseteq p^{\perp \perp}$. 
 If ${\cal C}_{p}$ is a 
connected component of ${\cal M}$ then ${\cal C}_{p}$ will be an atom. 
If this is not the case, let $q$ be a point in the border of ${\cal C 
}_{p}$. The point $q$ can only be a vertex or a point in ${\cal M}-{\cal B}$. 
Therefore if $p^{\perp \perp}$ contains a point in the border of ${\cal C}_p$ 
it will contain an atom. But any point in the border of ${\cal C}_p$ 
that can be reached by a 
time-like curve passing through a point in $p^{\perp \perp}$ must belong to 
$p^{\perp \perp}$ because of Lemma 1. Thus, if there are no point or bridge 
atoms in $p^{\perp \perp}$ we have 
${\cal C}_p=I^{+}({\cal C}_p)\cup I^{-}({\cal C}_p)$. 
A point in the set $-{\cal C}_p$ can not be joined by a time-like curve to 
${\cal C}_p$, and, as every point is in the future or the past of a different 
 point, 
it is  $-{\cal C}_p=I^{+}(-{\cal C}_p)\cup I^{-}(-{\cal C}_p)$. Therefore 
${\cal C}_p$ and $-{\cal C}_p$ are open disjoint sets that cover ${\cal M}$, 
and 
 ${\cal C}_{p}$ is a connected component of ${\cal M}$. 
 
 Resuming, we have shown 
 
\bigskip 
 
{\bf Theorem 2}: ${\cal L}({\cal M})$ is an atomic lattice. Its atoms are 
all the points in ${\cal M-B}$, the bridge atoms, and the connected 
components of ${\cal M}$ included in ${\cal C}$ 
 
\bigskip 
 
\smallskip Taking into account that ${\cal L}({\cal M})$ is orthomodular we 
also have, 
 
\bigskip 
 
{\bf Corollary: }${\cal L}({\cal M})$ is atomistic 
 
\bigskip 
 
That is, every element is the join of the atoms that includes. Thus, from 
the algebraic point of view all the bad points of the space-time disappear 
from the algebra, except the discrete bridge atoms and connected 
components included in ${\cal C}$ (see fig. 7). These later obviously 
 commute with all ${\cal L}({\cal M})$ and then belong to the centre. We can 
see that the same is true for the bridge atoms as follows. 
 
\begin{figure}[t] 
\centering 
\leavevmode 
\epsfysize=5cm \epsfbox{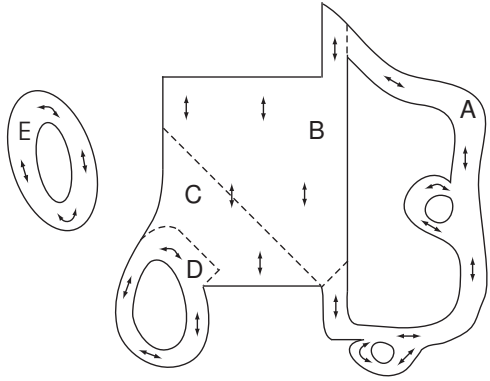} 
\smallskip 
\caption{A two dimensional space-time showing the different kinds of 
atoms. The direction of the light cones is marked with double arrows. The 
set A is a bridge atom. The set B is formed by vertex points that 
generates a set containing the bridge atom. 
The closed set C is equal to ${\cal M-B}$, and all points 
in this set are atoms. The sets D and E are formed by closed 
time-like curves. While E is a connected component of ${\cal M}$, and 
hence an atom, D is not. The lattice ${\cal L}({\cal M})$ is generated by the 
points in C and the isolated atoms A and E. These later commute with all 
other elements.  Note, however, that the lattice generated by ${\cal 
M-B}$ can not be deduced in general with the only knowledge of the 
space-time metric in ${\cal M-B}$.} \end{figure} 
 
\bigskip 
 
{\bf Lemma 5:} the bridge atoms belong to the centre of ${\cal L}({\cal M})$ 
 
\smallskip 
 
If an atom does not belong to $\beta ^{\perp }$ for a bridge $\beta $ it 
must be connected with $\beta $. If the atom is a point $p$ then as $p$ is 
connected with $\beta $ it implies that $p$ is a vertex, what is not 
consistent with $p$ being an atom. Then let the bridge $\chi $ be connected 
with $\beta $, it is easy to see that both will generate the same set, $ 
\beta ^{\perp \perp }=\chi ^{\perp \perp }$ (they form a unique bridge curve). 
 Thus an atom is either equal to 
$\beta ^{\perp \perp }$ or is included in $\beta ^{\perp }$, and therefore 
commutes with $\beta ^{\perp \perp }$. As ${\cal L}({\cal M})$ is atomistic, 
and every element is the join of its atoms, $ \beta ^{\perp \perp }$ will 
commute with all the elements of the lattice \cite{orthomodular}. 
 
\bigskip 
 
We have shown that if the manifold is not connected or if it contains at 
least two atoms and a bridge then ${\cal L}({\cal M})$ is reducible. We will 
complete the following 
 
\bigskip 
 
{\bf Theorem 3: }${\cal L}({\cal M})$ is trivial (equal to $\{O,{\cal M}\}$) 
if and only if ${\cal M}=\beta ^{\perp \perp }$ for a bridge $\beta $ or $ 
{\cal M}$ is connected and ${\cal M}={\cal C}$. If non trivial, ${\cal L}( 
{\cal M})$ is irreducible if and only if ${\cal M}$ is connected and does 
not contain any bridge curves 
 
\smallskip 
 
Assume that ${\cal M}$ does not have bridge curves, nor connected components 
includes in $ {\cal C}$. Then, its only atoms are the points in ${\cal M-B}$. 
We can identify all the sets in ${\cal L}({\cal M})$ with the set of its 
atoms, $ \hat{S}=S\cap ({\cal M-B})$. Let now $S$ be a proper element in 
${\cal L}( {\cal M})$. If there is a point $p$ in ${\hat {\cal 
M}}-\hat{S}-\hat{S} ^{\perp }$ it will be an atom that does not commute with 
$S$ (see eq.(\ref {commu})). Thus, if $S$ is in the centre of ${\cal L}({\cal 
M})$, it must be $ {\hat {\cal M}}=\hat{S}\cup \hat{S}^{\perp }$, and 
$\hat{S}\cap \hat{S} ^{\perp }=\emptyset $. Let $q$ be a point in ${\cal M}$, 
then, as the atoms that generate $q^{\perp \perp }$ are in $S$ or in 
$S^{\perp }$, we have $ q^{\perp \perp }=(q^{\perp \perp }\cap S)\vee 
(q^{\perp \perp }\cap S^{\perp })=(q^{\perp \perp }\cap S)\cup (q^{\perp 
\perp }\cap S^{\perp })$, the last equality coming from the form of the join 
for orthogonal sets. Thus, $q$ belongs to $S\cup S^{\perp }$, $S\cup S^{\perp 
}={\cal M}$, $S\cap S^{\perp }=O$. Then, all future and past of all points of 
$S$ must be in $S$, otherwise $S$ would be connected with $S^{\perp }$, and, 
as for a point $r$ $ \in S$ there is always a different point $s$ in a 
neighborhood that is in $ I^{+}(r)\cup I^{-}(r)$, and so $r$ $\in 
 I^{+}(s)\cup I^{-}(s)$ with $s \in S$.  Therefore 
$S=I^{+}(S)\cup I^{-}(S)$ is an open set. The same can be said for $ S^{\perp 
}$, $S^{\perp }=I^{+}(S^{\perp })\cup I^{-}(S^{\perp })$.  Therefore, ${\cal 
M}$ is non connected. 
 
\bigskip 
 
The case where ${\cal L}({\cal M})$ is nontrivial and irreducible 
is 
the most interesting for physical space-times. In that case there is a non 
countably number of atoms formed by the points in $\overline{{\cal M-B}}$, 
that can be consider the relevant space-time from the algebraic point of view. 
 
\section{Discussion} 
We have already mentioned that the lattice ${\cal C}({\cal H})$ of closed 
linear subspaces of the Hilbert space is complete, atomic, irreducible and 
orthomodular. It 
has a further property called the {\sl covering law}. An element $A$ in a 
lattice is 
said to {\sl cover} $B$ if $B\subseteq A$ and if for every element $C$ such 
that 
$B\subseteq C \subseteq A$ it is $C=A$ or $C=B$. The covering law means that 
given an atom $X$ and an element $A$ with $A\wedge X=O$, then $X\vee A$ covers 
$A$. There exist reconstruction theorems stating that 
a complete, atomic, irreducible, orthomodular lattice, satisfying the 
covering law, can be represented as a lattice of closed linear subspaces of a 
vector space with a Hermitian form \cite{beltram}. Under minor 
assumptions regarding the field of scalars for the vector space it is the 
 lattice of a 
Hilbert space. Thus, a lattice with this set of properties characterize 
almost uniquely Hilbert spaces.

\begin{figure}[tb] 
\centering 
\leavevmode 
\epsfysize=4cm 
\epsfbox{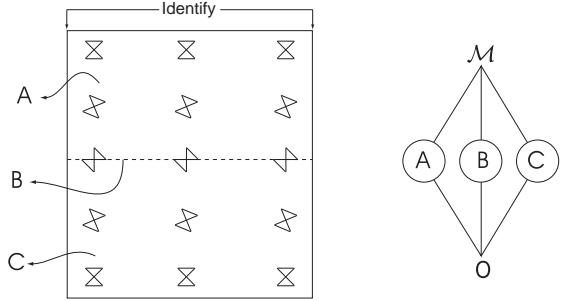} 
\smallskip 
\caption{The picture on the left shows a space-time constructed over a 
cylinder with no closed time-like curves but that is not strongly causal. 
This means that there is a point (all points in $B$) such that each of its 
neighborhoods is intersected by a time-like curve more than once. All points 
are atoms. The structure of the lattice ${\cal L}({\cal M})$ is shown on the 
 right. It is the horizontal sum of the lattices 
generated by the atoms in the sets $A$, $B$ and $C$. The lattice generated 
by $B$ is Boolean. The join of an atom in $A$ with an atom in $C$ (or in $B$) 
is the whole space-time, since all points are either in the future of a point 
in $C$ or in the past of a point in $A$.} \end{figure} 
 
The lattice ${\cal L}({\cal M})$ is also complete, atomic and orthomodular, 
and, in most of the physically interesting cases, irreducible. 
However, the covering law does not apply. For example, the join of two 
points connected by a time-like curve in Minkowski space does not cover any 
of them. This is the essential difference with ${\cal C}({\cal H})$, that 
obstructs the immersion of the lattice in a linear structure 
\cite{beltram,orthomodular}. 
 
It would be very interesting to explore what additional algebraic properties 
does the lattice ${\cal L}({\cal M})$ have, and, if possible, to find a set 
of properties that 
characterize the lattices of causally closed sets in the sense of a 
reconstruction theorem as the already mentioned. Such a possibility is 
suggested by the fact that the causal structure for stably causal space-times 
determines not only its conformal structure, but also the topological and 
differential structure of the manifold \cite{haw,haw2}. 
 
The next step in the construction of a dictionary between the causal structure 
and the algebraic setting is to translate into the lattice theoretical 
language the different causality conditions 
 such as strong causality and stable causality. Taub-NUT like space-times or 
the space-time of fig. 8 are not strongly causal and the structure of the 
resulting lattices is that of a {\sl horizontal sum} of several lattices. 
Given two 
lattices ${\cal L}_1$ and ${\cal L}_2$, their horizontal sum is the lattice 
formed by the set union of ${\cal L}_1$ and ${\cal L}_2$, where the unit and 
zero elements are identified (see fig. 8), so that the meet and join of an 
element 
of ${\cal L}_1$ with an element of ${\cal L}_2$ are $O$ and $I$ respectively. 
As a different kind of example we note that the Galilean space-time gives 
place to a 
causal lattice formed by the 
horizontal sum of the Boolean algebras corresponding to the space at 
different times. 
All this suggest that the  appearance of horizontal sums is characteristic of 
 non causally well behaved space-times. We postpone the general analysis for a 
future paper \cite{prep}. 
 
Another application of the lattice framework could be found in 
the study of the 
 asymptotic infinity. The spatial corner have a very simple expression in 
the algebraic context, being simply the image of ${\cal L}({\cal M})$ under 
an endomorphism of orthomodular lattices \cite{prep}. 
 
The causal structure is more fundamental in a logical sense than other 
aspects of 
space-time such as the metric. As we have already recalled, it also 
determines the 
topological and differential structure in well behaved space-times. 
In part motivated by this fact, there are in the literature approaches to 
quantum gravity that use a poset or lattice
 structure representing a causal order in space-time as a fundamental object. 
One is based on an axiomatic generalization of the history approach to 
quantum mechanics to a situation where the time evolution is less rigid than in
 ordinary quantum theory \cite{isham}. Another one,
 parts from discrete posets that would represent 
the causal structure at the Planck scale, and looks at the possible 
dynamics 
and the large scale structure that emerges \cite{sorkin}. It would be very 
interesting to see if the discussion in this paper, also focused in a 
lattice structure coming from the causal order, could add to these 
various invertigations. 
 
\section{Acknowledgments} 
I would like to thank C. Rovelli for useful discussions. This work was 
supported by CONICET, Argentina.

\end{document}